\documentclass[12pt]{article}
\usepackage{amsmath}
\usepackage{amssymb}
\usepackage{graphicx}
\usepackage{axodraw}
\usepackage{epsfig}
\usepackage{url}
\usepackage{fancybox}
\usepackage[small]{caption}
\begin{document}
\baselineskip 0.6cm

\def\simgt{\mathrel{\lower2.5pt\vbox{\lineskip=0pt\baselineskip=0pt
           \hbox{$>$}\hbox{$\sim$}}}}
\def\simlt{\mathrel{\lower2.5pt\vbox{\lineskip=0pt\baselineskip=0pt
           \hbox{$<$}\hbox{$\sim$}}}}
\def\simprop{\mathrel{\lower3.0pt\vbox{\lineskip=1.0pt\baselineskip=0pt
             \hbox{$\propto$}\hbox{$\sim$}}}}
\def\bra#1{\left< #1 \right|}
\def\ket#1{\left| #1 \right>}
\def\inner#1#2{\left< #1 | #2 \right>}

\begin{titlepage}

\begin{flushright}
UCB-PTH-12/07 \\
\end{flushright}

\vskip 1.3cm

\begin{center}
{\Large \bf The Static Quantum Multiverse}

\vskip 0.7cm

{\large Yasunori Nomura}

\vskip 0.4cm

{\it Berkeley Center for Theoretical Physics, Department of Physics,\\
 University of California, Berkeley, CA 94720, USA}

\vskip 0.1cm

{\it Theoretical Physics Group, Lawrence Berkeley National Laboratory,
 CA 94720, USA}

\vskip 0.8cm

\abstract{We consider the multiverse in the intrinsically quantum mechanical 
 framework recently proposed in Refs.~\cite{Nomura:2011dt,Nomura:2011rb}. 
 By requiring that the principles of quantum mechanics are universally 
 valid and that physical predictions do not depend on the reference frame 
 one chooses to describe the multiverse, we find that the multiverse 
 state must be static---in particular, the multiverse does not have a 
 beginning or end.  We argue that, despite its naive appearance, this 
 does not contradict observation, including the fact that we observe that 
 time flows in a definite direction.  Selecting the multiverse state 
 is ultimately boiled down to finding normalizable solutions to certain 
 zero-eigenvalue equations, analogous to the case of the hydrogen atom. 
 Unambiguous physical predictions would then follow, according to the 
 rules of quantum mechanics.}

\newpage

\tableofcontents

\end{center}
\end{titlepage}

\section{Introduction}
\label{sec:intro}

The goal of fundamental physics is to find a prescription in which 
(potentially) testable predictions can be made and, through it, to learn 
how nature works at the most fundamental level.  In the present way 
physics is formulated, this can be done in three steps:
\begin{itemize}
\item[(i)]
{\it ``Theory''} --- We must specify the fundamental structure of the 
theory, which consists of the following two parts:
\begin{itemize}
\item[(i-1)]
{\it Kinematics} --- We must understand what is a ``state'' which somehow 
represents the status of a physical system.  We must also understand 
how it is related to the observed reality.  For example, in conventional 
quantum mechanics a state is a ray in Hilbert space, which is related 
to reality through the Born rule, while in classical mechanics a state 
is a point in classical phase space (so is directly observable).
\item[(i-2)]
{\it Dynamics} --- We must know a (set of) fundamental law(s) the 
states obey.  In quantum mechanics it is the Schr\"{o}dinger equation, 
$i \frac{d}{dt} \ket{\Psi(t)} = H \ket{\Psi(t)}$, while in classical 
mechanics it is the Newton equation, $m\, \ddot{\bf x}(t) = {\bf F}$.
\end{itemize}
\item[(ii)]
{\it ``System''} --- We then need to specify a system we consider, which 
again  consists of two parts:
\begin{itemize}
\item[(ii-1)]
{\it Kinematics} --- We need to know the kinematical structure of the 
system.  In quantum mechanics this corresponds to specifying the Hilbert 
space, which is characterized by its dimension and operators acting on 
its elements.  In classical mechanics, it is given by the dimension of 
the phase space.
\item[(ii-2)]
{\it Dynamics} --- We also need to specify dynamics of the system.  In 
the examples in (i-2), we need to give the forms of $H$ and ${\bf F}$, 
respectively.
\end{itemize}
\item[(iii)]
{\it ``Selection Conditions''} --- Even if (i) and (ii) are known, we 
still need to provide ``selection conditions'' on a state.  Usually, 
they are given in the form of boundary conditions, for example as the 
knowledge one already has, e.g.\ $\ket{\Psi(0)}$ and $\{ {\bf x}(0), 
\dot{\bf x}(0) \}$, before making predictions on something unknown, 
e.g.\ $\ket{\Psi(t)}$ and $\{ {\bf x}(t), \dot{\bf x}(t) \}$ for $t > 0$.
\end{itemize}

To understand the ultimate structure of nature, we would want to do 
the above in the context of cosmology, and see whether the resulting 
predictions are consistent with what we observe.  In this respect, 
physics of eternal inflation---which occurs under rather general 
circumstances~\cite{Guth:1982pn}---has caused tremendous confusions 
in recent years.  A major problem has been the so-called measure problem:\ 
even if we know the initial state and its subsequent evolution, we 
cannot define (even probabilistic) predictions unambiguously.%
\footnote{There are several varying, though related, definitions of 
 the measure problem in literature.  In this paper we adopt the definition 
 as stated here.}
This occurs because in eternal inflation anything that can happen 
will happen infinitely many times, so it apparently leads to 
arbitrariness in predictions, associated with how these infinities 
are regularized~\cite{Guth:2000ka}.  Such an arbitrariness would prevent 
us from making well-defined predictions, so it seemed that to define the 
theory we needed to specify the exact way of regulating spacetime, where 
the infinities occur.  This would be quite uncomfortable, since then 
the theory requires a specification of a (ad hoc) regularization 
prescription {\it beyond the basic principles of quantum mechanics 
and relativity}.

Recently, a framework that addresses this problem has been proposed 
in Refs.~\cite{Nomura:2011dt,Nomura:2011rb}, which allows for an 
intrinsically quantum mechanical treatment of the eternally inflating 
multiverse (see Ref.~\cite{Nomura-review} for a review directed to 
a wide audience).  In this framework, physics is described in {\it a 
fixed reference (local Lorentz) frame} associated with a fixed reference 
point $p$, with spacetime existing only within its (stretched) apparent 
horizon.  An essential point is that the principles of quantum mechanics 
constrain the space of states ${\cal H}_{\rm QG}$~\cite{Nomura:2011rb} 
in such a way that the problem of infinity does not arise.  Namely, 
the correct identification of (ii-1) avoids the problem, without 
changing (i) from that of usual unitary quantum mechanics.  A state 
representing the multiverse $\ket{\Psi(t)}$ ``evolves'' deterministically 
and unitarily in ${\cal H}_{\rm QG}$, following the laws of quantum 
mechanics:\ $i \frac{d}{dt} \ket{\Psi(t)} = H \ket{\Psi(t)}$. 
Here, $t$ is an auxiliary parameter introduced to describe the 
``evolution'' of the state, and need not be directly related to 
physical time we observe.  Once the state $\ket{\Psi(t)}$ is known, 
physical predictions can be obtained through the (extended) Born 
rule~\cite{Nomura:2011dt,Nomura:2011rb} without suffering from an 
infinity or ambiguity.  This framework makes it possible that once 
a boundary condition on the state, e.g.\ $\ket{\Psi(t)}$ at some 
$t = t_0$, is given (element (iii)) and the explicit form of $H$ 
acting on ${\cal H}_{\rm QG}$ is understood, e.g.\ by studying string 
theory (element (ii-2)), then unambiguous predictions are obtained 
for any physical questions one asks.  While the framework does not 
achieve all of (i)--(iii), it does eliminate the ambiguity associated 
with the measure problem and provides a setting in which the remaining 
issues can be discussed.

In this paper we consider the issue of (iii) in the quantum mechanical 
framework of the multiverse described above.  We take the following 
hypothesis:
\begin{equation*}
\ovalbox{Hypothesis I: The laws of quantum mechanics are not violated.}
\label{eq:hypo-1}
\end{equation*}
This---in particular the fact that the evolution of a quantum state is 
deterministic and unitary---implies that the multiverse state exists 
all the way from $t = -\infty$ to $+\infty$.  Namely, the multiverse 
does not have a beginning or end.  (For recent discussions on 
the beginning of the eternally inflating multiverse, see, e.g., 
Refs.~\cite{Mithani:2012ii}.)  There are three potential issues 
in this picture:
\begin{itemize}
\item
{\it Uniqueness} --- What is the selection condition imposed on the 
multiverse state, on which physical predictions will depend?  In 
particular, what is the principle determining it?
\item
{\it Well-definedness} --- The (extended) Born rule formula in general 
involves $t$ integrals, which would now run from $-\infty$ to $+\infty$. 
Will this give well-defined probabilities?
\item
{\it Consistency} --- Are the resulting predictions consistent with 
observation?  In particular, are they consistent with the observed 
arrow of time, even if there is no beginning or end?
\end{itemize}

In this paper we argue that consistency with observation excludes 
the possibility that the selection condition is determined purely in 
${\cal H}_{\rm QG}$, without referring to an operator algebra.  In 
particular, this excludes the possibility that the multiverse is in 
the maximally mixed state in ${\cal H}_{\rm QG}$.  We then propose 
that the multiverse state must satisfy the following simple criterion:
\begin{equation*}
\ovalbox{Hypothesis II: Physical predictions do not depend on the 
 reference frame one chooses.}
\label{eq:hypo-2}
\end{equation*}
We show that this requirement leads to the condition
\begin{equation}
  \frac{d}{dt}\ket{\Psi(t)} = 0
\qquad\Leftrightarrow\qquad
  H \ket{\Psi(t)} = 0,
\label{eq:static}
\end{equation}
where we have taken $t$ to be the proper time at $p$; namely, we find 
that {\it the multiverse state must be static!}  We will argue that 
despite its naive appearance, this does not contradict observation, 
including the fact that we observe that time flows in a definite 
direction.  It simply gives constraints on the structure of $H$, on 
which we will allow for making arbitrary assumptions, given that its 
explicit form is not available under current theoretical technology. 
We will also argue that the hypothesis leads to unique and well-defined 
predictions for any physical questions, once one knows the explicit 
form of $H$ (element (ii-2) listed at the beginning).  Specifically, 
any physical question can be phrased in the form:\ given what we know 
$A$ about a state, what is the probability for it to be consistent 
also with $B$?  And the relevant probability is given by
\begin{equation}
  P(B|A) = \frac{\bra{\Psi} {\cal O}_{A \cap B} \ket{\Psi}}
    {\bra{\Psi} {\cal O}_A \ket{\Psi}},
\label{eq:prob-final}
\end{equation}
where $\ket{\Psi} \equiv \ket{\Psi(0)}$, and ${\cal O}_X$ is the 
operator projecting onto states consistent with condition $X$.

There are two comments.  First, given Hypothesis~I, Hypothesis~II arises 
as a consequence of general covariance (and its suitable extension 
to the quantum regime) if we assume that the multiverse is in a 
zero-eigenvalue eigenstate of global energy and boost operators.  This 
condition, therefore, provides another, more technical way of stating 
Hypothesis~II.  Second, without knowledge of the ultimate structure of 
$H$ in quantum gravity, the scenario presented here is not the only option 
available within the framework of Refs.~\cite{Nomura:2011dt,Nomura:2011rb}, 
although it seems to be the most natural possibility.  For example, 
one might imagine that the multiverse has a ``beginning,'' and evolves 
only thereafter.  (This violates both Hypotheses~I and II.)  The 
framework of Refs.~\cite{Nomura:2011dt,Nomura:2011rb} itself may 
still be applied in such a case.

The organization of this paper is as follows.  In the next 
section, we review the framework of the quantum multiverse given 
in Refs.~\cite{Nomura:2011dt,Nomura:2011rb}, and discuss the issue 
of selection conditions in that context.  In Section~\ref{sec:obs}, 
we reconsider what the arrow of time is.  We emphasize that the observed 
flow of time does not necessarily mean that the state is actually 
evolving.  In Section~\ref{sec:max-mixed}, we explore the possibility 
that the selection condition is expressed in ${\cal H}_{\rm QG}$ 
without referring to any quantum operator.  We find that this forces 
the multiverse to be in the maximally mixed state in ${\cal H}_{\rm QG}$, 
which is observationally excluded.  In Section~\ref{sec:stat}, we present 
our main scenario in which the multiverse state is determined by the two 
hypotheses described above.  We find this implies that the multiverse 
state must be static, and discuss how it can be realized in the 
cosmological context.  We also see that the scenario arises as a 
consequence of quantum mechanics and general covariance if we assume 
that the multiverse is in a zero-eigenvalue eigenstate of global 
energy and boost operators.  In Section~\ref{sec:consistency}, we 
discuss the consistency of the scenario with observation, specifically 
the observed arrow of time.  In Section~\ref{sec:discuss}, we provide 
our final discussions.  We draw a close analogy of the present scenario 
with the case of the hydrogen atom, underscoring the intrinsically 
quantum nature of the scenario.

\section{Framework---the Quantum Multiverse}
\label{sec:framework}

In this section we review the framework of 
Refs.~\cite{Nomura:2011dt,Nomura:2011rb}, describing the quantum 
multiverse.  We also discuss the issue of selection conditions in 
making predictions within this framework.

\subsection{The Hilbert space}
\label{subsec:Hilbert}

The framework is based on the principles of quantum mechanics.  In 
particular, we formulate it using Hamiltonian (canonical) quantum 
mechanics, although the equivalent Lagrangian (path integral) 
formulation should also be possible.  We take the Schr\"{o}dinger 
picture throughout.

Recall that to do Hamiltonian quantum mechanics, we need to fix all 
gauge redundancies.  Since these redundancies include coordinate 
transformations in a theory with gravity, states must be defined 
{\it as viewed from a fixed (local Lorentz) reference frame} associated 
with a fixed reference point $p$.  Moreover, to avoid violation of the 
principles of quantum mechanics, they must represent only spacetime 
regions within the (stretched) apparent horizons of $p$, as suggested 
first in the study of black hole physics~\cite{Susskind:1993if}. 
Together with the states associated with spacetime singularities, 
these states form the Hilbert space for quantum gravity 
${\cal H}_{\rm QG}$.

The construction of ${\cal H}_{\rm QG}$ can proceed analogously to 
the usual Fock space construction in quantum field theory.  For a set 
of fixed semi-classical geometries ${\cal M} = \{ {\cal M}_i \}$ having 
the same apparent horizon $\partial {\cal M}$, the Hilbert space is 
given by
\begin{equation}
  {\cal H}_{\cal M} = {\cal H}_{{\cal M}, {\rm bulk}} 
    \otimes {\cal H}_{{\cal M}, {\rm horizon}},
\label{eq:ST-H_M}
\end{equation}
where ${\cal H}_{{\cal M}, {\rm bulk}}$ and ${\cal H}_{{\cal M}, {\rm 
horizon}}$ represent Hilbert space factors associated with the degrees 
of freedom inside and on the horizon $\partial {\cal M}$.  The dimensions 
of these factors are both $\exp({\cal A}_{\partial {\cal M}}/4)$, where 
${\cal A}_{\partial {\cal M}}$ is the area of the horizon in Planck units:
\begin{equation}
  {\rm dim}\,{\cal H}_{\cal M}
  = {\rm dim}\,{\cal H}_{{\cal M}, {\rm bulk}} \times 
    {\rm dim}\,{\cal H}_{{\cal M}, {\rm horizon}} 
  = \exp\left(\frac{{\cal A}_{\partial {\cal M}}}{2}\right),
\label{eq:H_M-dimension}
\end{equation}
consistently with the holographic principle~\cite{'tHooft:1993gx}. 
The full Hilbert space for dynamical spacetime is then given by the 
direct sum of the Hilbert spaces for different ${\cal M}$'s
\begin{equation}
  {\cal H} = \bigoplus_{\cal M} {\cal H}_{\cal M}.
\label{eq:ST-H}
\end{equation}
In addition, the complete Hilbert space for quantum gravity must contain 
``intrinsically quantum mechanical'' states, associated with spacetime 
singularities~\cite{Nomura:2011rb}:
\begin{equation}
  {\cal H}_{\rm QG} = {\cal H} \oplus {\cal H}_{\rm sing},
\label{eq:QG-H}
\end{equation}
where ${\cal H}_{\rm sing}$ represents the Hilbert space for the 
singularity states.  The evolution of the multiverse state $\ket{\Psi(t)}$, 
which represents the entire multiverse, is deterministic and unitary 
in ${\cal H}_{\rm QG}$, but not in ${\cal H}_{\cal M}$ or ${\cal H}$.

The dimension of the complete Hilbert space ${\cal H}_{\rm QG}$ is 
infinite, as the dimensions of Hilbert subspaces associated with stable 
Minkowski space and spacetime singularities are infinite:
\begin{equation}
  {\rm dim}\,{\cal H}_{\rm Minkowski} = \infty,
\qquad
    {\rm dim}\,{\cal H}_{\rm sing} = \infty.
\label{eq:dim-inf}
\end{equation}
This implies, by the second law of thermodynamics, that a {\it generic} 
multiverse state in ${\cal H}_{\rm QG}$ will evolve at large $t$ into 
a superposition of terms corresponding to supersymmetric Minkowski space 
or spacetime singularity:
\begin{equation}
  \ket{\Psi(t)}
  \,\,\stackrel{t \rightarrow \infty}{\longrightarrow}\,\, 
    \sum_i a_i(t) \ket{\mbox{supersymmetric Minkowski space $i$}}
  \,+\, \sum_j b_j(t) \ket{\mbox{singularity state $j$}},
\label{eq:asympt}
\end{equation}
where we have assumed that the only absolutely stable Minkowski 
vacua are supersymmetric ones, as suggested by the string landscape 
picture~\cite{Bousso:2000xa}.

Note that an infinite number of states exist only in a Hilbert subspace 
associated with a spacetime singularity or a Minkowski space {\it in which 
the area of the apparent horizon diverges ${\cal A}_{\partial {\cal M}} 
= \infty$}.  In particular, the number of states associated with a 
fixed Friedmann-Robertson-Walker (FRW) time in a Minkowski bubble is 
finite for any finite energy density $\rho$, since the area of the 
apparent horizon is given by ${\cal A}_{\partial {\cal M}} = 3/2\rho$ 
(with $\rho$ in Planck units)~\cite{Bousso:2002ju-FRW}, so that 
${\cal A}_{\partial {\cal M}} < \infty$ for $\rho > 0$.

\subsection{The (extended) Born rule}
\label{subsec:Born}

For a given multiverse state $\ket{\Psi(t)}$, physical predictions can 
be obtained following the rules of quantum mechanics.  An important point 
is that the ``time'' parameter $t$ here is simply an auxiliary parameter 
introduced to describe the ``evolution'' of the state.  The physical 
information is only in {\it correlations} between events; specifically, 
time evolution of a physical quantity $X$ is nothing more than a 
correlation between $X$ and a quantity that can play the role of time, 
such as the location of the hands of a clock or the average temperature 
of the cosmic microwave background in our universe.  A particularly 
useful choice for $t$ is the proper time at $p$, which we will assume 
for the rest of the paper.

Any physical question can then be phrased as:\ given what we know $A$ 
about a state, what is the probability for that state to be consistent 
also with condition $B$?  In the context of the multiverse, this 
probability is given by~\cite{Nomura:2011dt}
\begin{equation}
  P(B|A) = \frac{\int\!dt \bra{\Psi(0)} U(0,t)\, 
      {\cal O}_{A \cap B}\, U(t,0) \ket{\Psi(0)}}
    {\int\!dt \bra{\Psi(0)} U(0,t)\, {\cal O}_A\, U(t,0) \ket{\Psi(0)}},
\label{eq:prob}
\end{equation}
where $U(t_1,t_2) = e^{-iH(t_1-t_2)}$ is the ``time evolution'' operator 
with $H$ being the Hamiltonian of the entire system for a fixed ``time'' 
parameterization $t$ (here the proper time at $p$), and ${\cal O}_X$ 
is the operator projecting onto states consistent with condition $X$. 
Note that since we have already fixed a reference frame, conditions $A$ 
and $B$ in general must involve specifications of ranges of location 
and velocity in which a physical object must be with respect to the 
reference point $p$.

As we will discuss in more detail in Section~\ref{sec:obs}, the formula 
in Eq.~(\ref{eq:prob}) can be used to answer any physical questions 
including those about dynamical evolution of a system, despite the 
fact that conditions $A$ and $B$ both act at the same moment $t$. 
We therefore base all our discussions on Eq.~(\ref{eq:prob}) in this 
paper.  (For a different formula that can be used more easily in many 
practical contexts, see Ref.~\cite{Nomura:2011rb}.)  The $t$ integrals 
in the equation run over the entire region under consideration.  Suppose, 
for example, that we know the universe/multiverse is in a particular, 
e.g.\ eternally inflating, state $\ket{\Psi(0)}$ at $t = 0$, and want 
to predict what happens in $t > 0$.  In this case, the integrals must 
be taken from $t = 0$ to $\infty$, since condition $A$ may be satisfied 
at any value of $t > 0$ in some component of $\ket{\Psi(t)}$.  Note 
that despite the integrals running to $\infty$ the resulting probability 
is well-defined, because Eq.~(\ref{eq:asympt}) prohibits an event from 
occurring infinitely many times with a finite probability, which would 
cause divergences.

\subsection{The issue of selection conditions}
\label{subsec:s-c}

What kind of predictions does the framework described above allow us 
to make?  While the framework addresses the issues of infinity and the 
ambiguity associated with it (i.e.\ the measure problem as defined here), 
it is certainly not complete.  In particular, ...
\begin{itemize}
\item[(a)]
{\it ``Unspecified System''} --- We did not identify the system {\it 
explicitly}.  Specifically, the complete theory of quantum gravity is 
not known, so that we do not know the form of $H$, especially the part 
acting on the horizon degrees of freedom.  This particular issue can be 
bypassed if we focus only on questions addressed at the semi-classical 
level.  Even then, however, current technology does not give us the 
explicit form of $H$, e.g.\ the structure of the string landscape.
\item[(b)]
{\it ``Selection Conditions''} --- Predictions in general depend on the 
selection condition we impose on $\ket{\Psi(t)}$ (even if we know $H$ 
explicitly).  For example, in the situation considered at the end of the 
previous subsection, they depend on the initial condition $\ket{\Psi(0)}$.
\end{itemize}

These limitations may still allow us to make certain predictions, possibly 
with some assumption on the dynamics of the system.  First of all, if 
we are interested in a system localized in a small region compared with 
the horizon scale, then we can make predictions on the evolution of the 
system (i.e.\ correlation with a physical quantity that plays the role 
of time) using prior information about the system---indeed, one can 
show that Eq.~(\ref{eq:prob}) is reduced to the standard Born rule in 
such a case.  Second, if we are interested in quantities whose distributions 
in $H$ are reasonably inferred in an anthropically allowed range, then 
we can predict the probability distribution of these quantities seen 
by a typical observer, under the assumption that the selection condition 
provides a statistically uniform prior~\cite{Weinberg:1987dv}.  This is, 
for example, the case if we are interested in the probability distribution 
of the cosmological constant one observes~\cite{Larsen:2011mi}.

However, if we want to answer general ``multiversal'' questions, e.g.\ 
if we want to predict the probability distribution of the structure 
of the low-energy Lagrangian found by an intellectual observer in 
the multiverse, then we would need to address both (a) and (b) above. 
(What the intellectual observer means can be specified explicitly by 
condition $A$.)  For (a), one could hope that future progress, e.g.\ 
in string theory, might provide us (at least the relevant information 
on) the form of $H$ in ${\cal H}_{\rm QG}$.  But what about (b)?

There are at least three aspects which make this problem substantial:
\begin{itemize}
\item
One might speculate that a physical theory only allows for relating a 
given initial state to another final state, which is indeed the case 
in conventional Newtonian and quantum mechanics.  In the present context, 
this implies that to make general predictions, we need to know the state 
$\ket{\Psi(t)}$ explicitly for some $t$.  This is, however, impossible 
to do observationally!  Quantum mechanics does not allow us to 
know the exact state {\it including us, the observer}.  Moreover, 
$\ket{\Psi(t)}$ is the quantum state for the whole multiverse, so 
it in general contains terms representing different semi-classical 
universes than what we live in.
\item
General predictions in the multiverse, therefore, will be possible 
only if we have a {\it theoretical} input on the selection condition 
of $\ket{\Psi(t)}$.  Suppose it takes the form of a specific ``initial 
condition,'' $\ket{\Psi(0)}$.  Then, the predictions depend on 
$\ket{\Psi(0)}$, so that, unless we have a separate theory of the 
initial condition, the uniqueness of (even statistical) predictions 
will be lost.
\item
Imagine that there is, indeed, a theory of the initial condition giving 
a particular state $\ket{\Psi(0)}$, and that the framework described 
in Sections~\ref{subsec:Hilbert} and \ref{subsec:Born} applies only 
to $t > 0$.  In this case, the laws of quantum mechanics, especially 
deterministic and unitary evolution of the state, is violated at $t = 0$. 
While this is possible, it would be more comfortable if fundamental 
principles, such as those of quantum mechanics, do not have an 
``exception'' like this.
\end{itemize}

In the rest of the paper, we will address the problem of selection 
conditions, i.e.\ issue (b), from the viewpoint of extrapolating the 
principles of quantum mechanics to the maximum extent possible.  By 
postulating a certain simple criterion, and requiring consistency with 
observation, we will arrive at the picture that the multiverse state 
must, in fact, be static.  This provides a strong selection of the 
possible states.  The observed flow of time arises from the structures 
of $H$ in ${\cal H}_{QG}$, and not because of a $t$ dependence of 
$\ket{\Psi(t)}$.

\section{The Observational ``Data''}
\label{sec:obs}

Any selection condition imposed on the multiverse state must not lead to 
results inconsistent with observation, if it is to do with nature.  The 
basic observational fact in our universe is that we see time flow in 
a definite direction, and predictions of a theory must not contradict 
it.  As we will see, this seemingly weak requirement, in fact, provides 
a powerful tool to determine the selection condition.  Here, we carefully 
consider what the observed flow of time actually means in the context 
of the quantum multiverse.

\subsection{What is the arrow of time?}
\label{subsec:arrow-time}

What does the fact that we see time flow really mean?  At the most 
elementary level, it just means that the memory state of my (or your) 
brain is consistent with the hypothesis that it is generated by an 
environment whose coarse-grained entropy evolves from lower to higher 
values.  The point is that the states consistent with such a hypothesis 
are very special ones among all the possible states the brain can take. 
What the fundamental theory must explain is why my brain is in one 
of these highly exceptional states.

\begin{figure}[t]
\begin{center}
  \includegraphics[width=16cm]{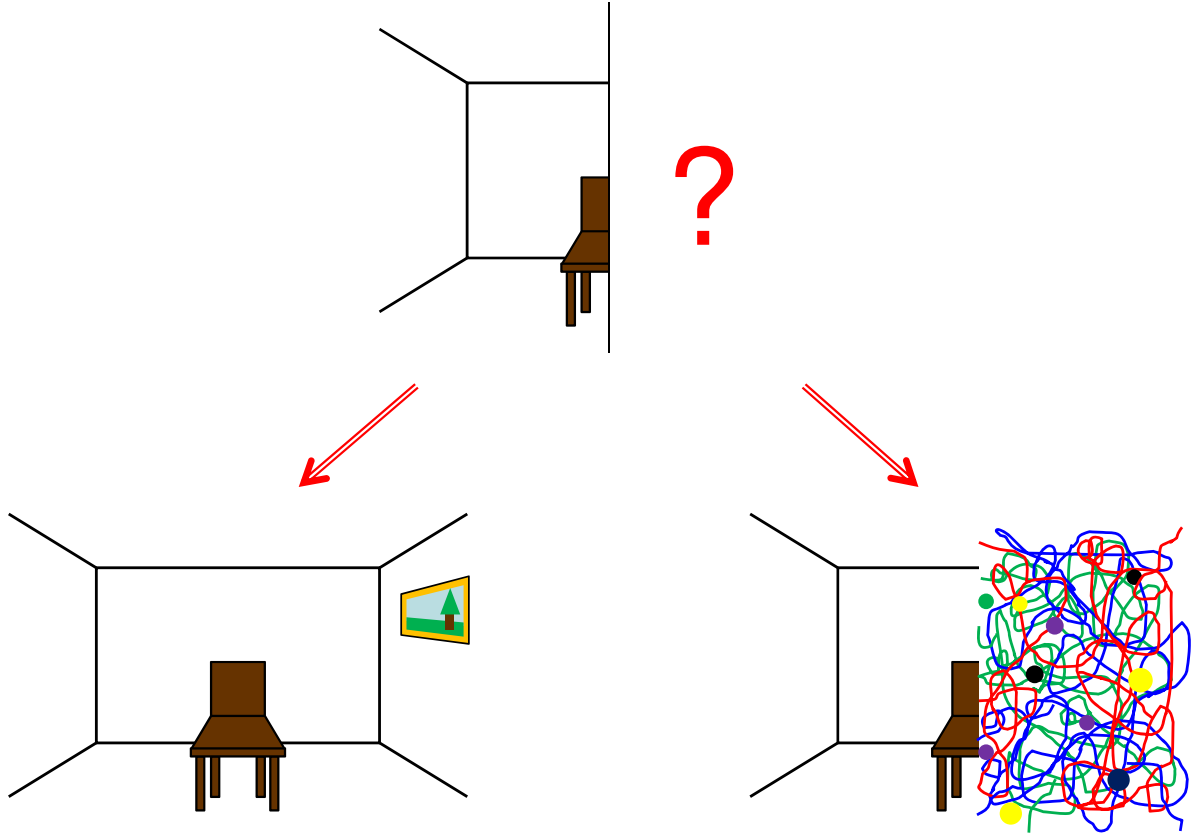}
\caption{Suppose you know that there are a half of a chair and of a 
 room in the first half of the scene (the upper picture).  In a regular 
 ordered world, you expect the second half of the scene contains the 
 other half of the chair and the room, possibly with some other things 
 (the lower left picture).  On the other hand, the number of such 
 states is much smaller than that of states in which the second half 
 contains random, disordered configurations (the lower right picture).}
\label{fig:chair}
\end{center}
\end{figure}
To illustrate the basic idea further, let us consider a more corporeal 
example of a chair in a room.  Suppose you are looking at only a half 
of the scene and find a half of a chair and of a room there; see the 
upper picture in Fig.~\ref{fig:chair}.  What would you expect to be 
in the other half?  In the ordered world we live in, we expect to see the 
other half of the chair and the room, possibly with some other things 
such as a painting on the wall, as depicted in the lower left picture 
in Fig.~\ref{fig:chair}.  However, any such configurations are extremely 
rare among all the possible configurations physically allowed and 
consistent with the first half of the scene.  The vast majority of 
these general configurations correspond to the ones in which the 
other half of the scene is completely disordered, as depicted in the 
lower right picture in Fig.~\ref{fig:chair}.  The arrow of time refers 
to the fact that we always find ordered configurations (as in the lower 
left picture) rather than disordered ones (as in the lower right picture) 
in any similar situation, i.e.\ not only for a chair in a room but 
also for other objects.  Such ordered configurations can be naturally 
expected if the entire system is evolved from a state having a much 
lower coarse-grained entropy; otherwise, we would expect disordered 
ones since the number of states corresponding to disordered configurations 
is much larger than that corresponding to ordered ones.

In the context of the multiverse, the fact that we live in our universe 
and see the arrow of time tells us two things:
\begin{itemize}
\item[(A)]
A typical observer among all the ``conscious'' observers  in the 
multiverse (including fluke, Boltzmann brain observers~\cite{Dyson:2002pf}) 
must live in a universe consistent with our current knowledge, i.e.\ 
a universe whose low energy physics is described by the standard model 
of particle physics and cosmology.
\item[(B)]
When we ask any conditional probability $P(B|A)$ within our universe, 
i.e.\ when precondition $A$ is chosen such that it selects a situation 
in our universe (e.g.\ my brain state), the answer should be dominated 
by one that arises from a low coarse-grained entropy state through 
evolution.
\end{itemize}
These two are the only things we definitely know from observation about 
the structure of the multiverse; for example, the arrow of time may 
not exist in other universes, i.e.\ the probabilities may be dominated 
by disordered configurations in those universes.  What we must require 
is that the theory must (at least) be compatible with these two 
conditions.

The above discussion shows that the following two statements are literally 
{\it equivalent} as concepts:\ ``An observer sees the arrow of time'' 
and ``There is no Boltzmann brain problem.''  This is consistent with 
the picture presented recently by Bousso~\cite{Bousso:2011aa}, who 
analyzed the arrow of time in the context of the evolving multiverse 
in the landscape.  Historically, the argument like the one here was 
first used to exclude the possibility that our universe, which has 
a positive cosmological constant, is absolutely stable~\cite{Dyson:2002pf}. 
It was also argued in Ref.~\cite{Nomura:2011rb} that it excludes the 
possibility that the multiverse is a closed, finite system if it has 
a {\it generic} initial condition in ${\cal H}_{\rm QG}$.  This possibility, 
however, is allowed if the selection condition imposed on the entire 
multiverse state is special, as is the case in the scenario considered 
in this paper.

In summary, a selection condition imposed on the multiverse state must 
be such that the resulting probabilities are consistent with conditions 
(A) and (B) listed above.  In particular, this leads to the following 
corollary:
\begin{itemize}
\item[($\ast$)]
Any selection condition on $\ket{\Psi(t)}$ that leads to an (almost) 
equal probability for all the possible states in ${\cal H}_{\rm QG}$ 
corresponding to our universe is observationally excluded.
\end{itemize}
This is because such a scenario would lead to the probabilities being 
dominated by disordered configurations in our universe, contradicting 
observation.  This condition will play an important role in rejecting 
a possible selection condition in Section~\ref{sec:max-mixed}.

\subsection{Is the multiverse really evolving?}
\label{subsec:evolve}

The consideration given above also illuminates the following question:\ 
is the multiverse really evolving?  The answer is:\ it need not.  In order 
to be consistent with the observed arrow of time, it is only necessary 
that the probabilities {\it in our universe} are dominated by configurations 
that are consistent with the hypothesis that the system has evolved 
from a lower coarse-grained entropy state.  This, however, does not 
necessarily mean that the multiverse state $\ket{\Psi(t)}$ is actually 
evolving in $t$.  It simply says that the probabilities obtained from 
$\ket{\Psi(t)}$ should be consistent with the hypothesis that our universe 
has evolved from a lower entropy state.

One might think that we actually ``witnessed'' that the state evolved 
as we came into being and grew.  The interpretation of this fact, 
however, needs care---all we know is that our memory states are such 
that they are {\it consistent with} those obtained by interacting with 
environments that evolve from lower to higher entropy states.  Similarly, 
we usually consider that our universe has evolved from the early big-bang, 
but all we really know is that the current state of the universe is 
consistent with the hypothesis that it has evolved from a lower entropy, 
big-bang state.  As we have seen in the previous subsection, what 
these observations are really telling us is that in our universe 
different parts of physical configurations are correlated in certain 
(very) special ways.  They do not mean that the multiverse state 
$\ket{\Psi(t)}$ {\it must} be evolving.

The question of whether a physical system is viewed as evolving or not, 
therefore, can be determined by asking questions about a ``current'' 
configuration, i.e.\ configuration at a fixed value of $t$.  If the 
configuration is consistent with the hypothesis that the system has 
evolved from a lower entropy state, then we {\it interpret} it as the 
system evolving---it is not necessary that the state itself is actually 
changing with $t$.  To do such a determination, it is enough to use 
the formula of Eq.~(\ref{eq:prob}), in which conditions $A$ and $B$ 
act at the same moment.  In fact, in quantum mechanics, when 
we obtain information about a system we do that indirectly by observing 
imprints in the environment left by the system~\cite{q-Darwinism}, 
so this is almost exactly what we do in reality when we study the 
``history'' of a system.

Summarizing, the observed flow of time does not require that the 
multiverse state is actually changing with $t$.  It simply requires 
that the resulting probabilities satisfy the two conditions described 
in the previous subsection:\ (A) and (B).  The probability formula 
in which conditions $A$ and $B$ both act at the same moment can be 
used to answer any physical questions, including those about a system 
that we interpret as dynamically evolving.

\section{Selection Conditions and Operators}
\label{sec:max-mixed}

We now start exploring possible selection conditions that can be imposed 
on the multiverse state.  As stated in the introduction, we consider 
that the laws of quantum mechanics are not violated (Hypothesis~I), 
which forces the multiverse state to exist for all values of $t$:\ from 
$-\infty$ to $+\infty$.  This implies that, once a selection condition 
is given at a particular moment, which we take as $t = 0$, then the 
state is uniquely determined by solving the Schr\"{o}dinger equation 
both forward and backward in $t$.

In this section, we ask the following question:\ can the selection 
condition be given in Hilbert space ${\cal H}_{\rm QG}$ without referring 
to any quantum operator?  If this is possible, then it would imply that 
the form of the selection condition, written purely in terms of quantum 
states, must be basis independent, since we cannot specify a basis without 
knowledge of operators and how they act on elements in the Hilbert space. 
(Note that Hilbert space itself does not contain any physical information 
except for its dimension, i.e.\ any complex Hilbert spaces having the 
same dimension are identical with each other.)  We will see that there 
is only one possible selection condition satisfying this criterion, 
and that it is observationally excluded.  We will therefore learn that 
the expression for the selection condition in ${\cal H}_{\rm QG}$ must 
involve some information about the quantum operators.

\subsection{The selection condition without an operator}
\label{subsec:state-MM}

Suppose that the multiverse is in a pure state, and that the selection 
condition at $t = 0$ is given by
\begin{equation}
  \ket{\Psi(0)} = \sum_i c_i \ket{\alpha_i},
\label{eq:bc-pure}
\end{equation}
where $\ket{\alpha_i}$ represents a complete, orthonormal basis for 
the elements in ${\cal H}_{\rm QG}$, and $c_i$ are fixed coefficients 
characterizing the selection condition.  Can the expression in 
Eq.~(\ref{eq:bc-pure})---{\it including the values of $c_i$}---be 
basis independent?

Consider that we perform an arbitrary basis change
\begin{equation}
  \ket{\alpha_i} = \sum_j U_{ij} \ket{\alpha'_j},
\label{eq:basis-change}
\end{equation}
where $U_{ij}$ is an arbitrary unitary matrix.  In the new basis, the 
expression in Eq.~(\ref{eq:bc-pure}) is written as $\ket{\Psi(0)} = 
\sum_i c'_i \ket{\alpha'_i}$, where the new coefficients $c'_i$ are 
given by $c'_i = \sum_j c_j U_{ji}$.  In order for the form of the 
selection condition to be basis independent, we need to have
\begin{equation}
  c_i = c'_i = \sum_j c_j U_{ji}
\label{eq:pure-indep}
\end{equation}
{\it for an arbitrary $U_{ij}$}.  This condition cannot be satisfied 
unless $c_i = 0$ for all $i$.  Therefore, it is not possible to write 
a selection condition without referring to any quantum operator if 
the multiverse state is pure.

Suppose now that the multiverse is in an intrinsically mixed state, 
which takes the form
\begin{equation}
  \rho(0) = \sum_{i,j} d_{ij} \ket{\alpha_i} \bra{\alpha_j}
\label{eq:bc-mixed}
\end{equation}
at $t = 0$, where $d_{ij}$ is a positive semi-definite Hermitian matrix. 
The basis change in Eq.~(\ref{eq:basis-change}) then leads to $\rho(0) 
= \sum_i d'_{ij} \ket{\alpha'_i} \bra{\alpha'_j}$, where the new 
coefficients are given by $d'_{ij} = \sum_{k,l} U_{ik} d_{kl} U^*_{jl}$. 
In order for the selection condition to be basis independent, we 
must have
\begin{equation}
  d_{ij} = d'_{ij} = \sum_{k,l} U_{ik} d_{kl} U^*_{jl}
\label{eq:mixed-indep}
\end{equation}
for an arbitrary $U_{ij}$.  This has the unique solution (up to the 
overall coefficient):
\begin{equation}
  d_{ij} \propto \delta_{ij}.
\label{eq:d-indep}
\end{equation}
We thus find that the requirement is satisfied if the multiverse state 
is specified by
\begin{equation}
  \rho(0) \propto \sum_i \ket{\alpha_i} \bra{\alpha_i},
\label{eq:rho-MM}
\end{equation}
namely if the multiverse is in the maximally mixed state in 
${\cal H}_{\rm QG}$ at $t = 0$.

\subsection{Can the multiverse be in the maximally mixed state?}
\label{subsec:max-ig}

Once the selection condition is given by Eq.~(\ref{eq:rho-MM}), the 
multiverse state $\rho(t)$ for arbitrary $t$ can be obtained using 
the evolution equation
\begin{equation}
  \rho(t) = U(t,0)\, \rho(0)\, U(0,t).
\label{eq:mixed-evol}
\end{equation}
Since $\rho(0)$ is proportional to the unit matrix in ${\cal H}_{\rm QG}$, 
however, this gives
\begin{equation}
  \rho(t) = \rho(0),
\label{eq:mixed-0}
\end{equation}
i.e.\ the multiverse is in the maximally mixed state at all times.

Equations~(\ref{eq:rho-MM}) and (\ref{eq:mixed-0}) imply that all the 
possible states in ${\cal H}_{\rm QG}$ corresponding to our universe are 
equally probable.  This is exactly the possibility that is observationally 
excluded by corollary~($\ast$) in Section~\ref{subsec:arrow-time}. 
Since we have arrived at this conclusion only by assuming that the 
selection condition is written without referring to a quantum operator 
in ${\cal H}_{\rm QG}$, we learn that the condition must in fact 
involve a quantum operator.  The significance of this result lies 
in the fact that in quantum mechanics, operators are the objects that 
contain information about the system---the condition imposed on the 
multiverse state must reflect the structure of the system.

\section{The Static Quantum Multiverse}
\label{sec:stat}

What operators can be used in the condition imposed on the multiverse 
state?  Since the multiverse contains many universes in which low energy 
physical laws differ, they cannot be ``vacuum specific'' operators. 
In this section, we identify candidate operators---those generating 
reference frame changes and that generating evolution.

We then impose the requirement that physical predictions are independent 
of a reference frame one chooses to describe the multiverse (Hypothesis~II 
in the introduction).  We will see that this implies that the multiverse 
state is independent of $t$, i.e.\ it must be static.  As discussed 
in Section~\ref{subsec:evolve}, this does not necessarily contradict 
observation.  (The consistency with the observed flow of time will 
be discussed further in Section~\ref{sec:consistency}.)  We will also 
see that with Hypothesis~I, Hypothesis~II can be viewed as a consequence 
of requiring that the multiverse is in an eigenstate of global energy 
and boost operators with zero eigenvalues.

\subsection{Reference frame changes}
\label{subsec:ref-change}

Recall that in the framework of Refs.~\cite{Nomura:2011dt,Nomura:2011rb}, 
quantum states allowing for spacetime interpretation, i.e.\ elements 
of ${\cal H} \subset {\cal H}_{\rm QG}$, represent only the spacetime 
regions inside and on the (stretched) apparent horizons as viewed from 
a fixed reference frame associated with a fixed reference point $p$. 
What happens if we change the reference frame?

Consider a state representing a configuration in de~Sitter space.  If 
we perform a spatial translation, which is equivalent to shifting the 
location of $p$, then it will necessarily mix the degrees of freedom 
inside and on the horizon because the state is defined only in the 
restricted spacetime region.  This is precisely the phenomenon we 
call the observer dependence of the horizon:\ (some of) the degrees 
of freedom associated with internal space for one observer are described 
as those associated with the horizon by another.  Next, consider 
a state which will later form a black hole, with $p$ staying outside 
of the black hole horizon.  Such a state will not contain the spacetime 
region inside the black hole horizon because it will be outside $p$'s 
horizon.  Now, imagine that we change the reference frame by performing 
a boost at an early time so that $p$ will be inside the black hole 
horizon at late times.  In this new frame, the state at late times 
{\it does} contain the spacetime region inside the black hole horizon, 
although now it does {\it not} contain Hawking radiation quanta escaping 
to the future null infinity, which were included in the state before 
performing the reference frame change.  This is exactly the phenomenon 
of black hole complementarity~\cite{Susskind:1993if}.  The present 
framework, therefore, allows us to understand the two phenomena 
described above in a unified manner as special cases of general 
reference frame changes~\cite{Nomura:2011rb}; in particular, the 
concept of spacetime depends on the reference frame.

As any symmetry transformation, reference frame changes must be 
represented by unitary transformations acting on Hilbert space 
${\cal H}_{\rm QG}$.  What is the set of generators representing 
these transformations, and what is the algebra they satisfy?

In the limit $G_N \rightarrow 0$, the set of transformations associated 
with the reference frame changes and a shift of the origin of $t$ (time 
translation) is reduced to the standard Poincar\'{e} transformations, 
which is analogous to the fact that the standard Poincar\'{e} 
group is reduced to the Galilean group in the limit $c \rightarrow 
\infty$~\cite{Nomura:2011rb}.  Here, $G_N$ and $c$ are Newton's 
constant and the speed of light, respectively.  In the case of the 
reduction associated with $c \rightarrow \infty$, the structure of 
infinitesimal transformations changes.  This is seen clearly in the 
Poincar\'{e} algebra:
\begin{equation}
\begin{array}{c}
  [ J_{[ij]}, J_{[kl]} ] 
  = i \left( \delta_{ik} J_{[jl]} - \delta_{il} J_{[jk]} 
    - \delta_{jk} J_{[il]} + \delta_{jl} J_{[ik]} \right),
\\[10pt]
  [ J_{[ij]}, K_k ] = i \left( \delta_{ik} K_j - \delta_{jk} K_i \right),
\qquad
  [ K_i, K_j ] = - \frac{i}{c^2} J_{[ij]},
\\[10pt]
  [ J_{[ij]}, P_k ] = i \left( \delta_{ik} P_j - \delta_{jk} P_i \right),
\qquad
  [ K_i, P_j ] = \frac{i}{c^2} \delta_{ij} H,
\qquad
  [ P_i, P_j ] = 0,
\\[10pt]
  [ J_{[ij]}, H ] = [ P_i, H ] = [ H, H ] = 0,
\qquad
  [ K_i, H ] = i P_i,
\end{array}
\label{eq:Poincare-alg}
\end{equation}
where $J_{[ij]}$, $K_i$, and $P_i$ are the generators of spatial rotations, 
boosts, and spatial translations, respectively, and we have exhibited $c$ 
explicitly.  This algebra is reduced to a different algebra, i.e.\ that 
of the Galilean group, as $c \rightarrow \infty$:
\begin{equation}
\begin{array}{c}
  [ J_{[ij]}, J_{[kl]} ] 
  = i \left( \delta_{ik} J_{[jl]} - \delta_{il} J_{[jk]} 
    - \delta_{jk} J_{[il]} + \delta_{jl} J_{[ik]} \right),
\\[10pt]
  [ J_{[ij]}, K_k ] = i \left( \delta_{ik} K_j - \delta_{jk} K_i \right),
\qquad
  [ K_i, K_j ] = 0,
\\[10pt]
  [ J_{[ij]}, P_k ] = i \left( \delta_{ik} P_j - \delta_{jk} P_i \right),
\qquad
  [ K_i, P_j ] = i \delta_{ij} M,
\qquad
  [ P_i, P_j ] = 0,
\\[10pt]
  [ J_{[ij]}, H ] = [ P_i, H ] = [ H, H ] = 0,
\qquad
  [ K_i, H ] = i P_i,
\end{array}
\label{eq:Galilean-alg}
\end{equation}
where we have rescaled $H \rightarrow c^2 M + H$ to allow for the 
possibility that the original $H$ has a constant piece that goes as 
$c^2$.  Can the algebra corresponding to the reference frame changes 
and time translation have extra terms beyond Eq.~(\ref{eq:Poincare-alg}) 
that disappears in the limit $G_N \rightarrow 0$?

One can immediately see that it cannot.  The generators of the reference 
frame changes consist of $J_{[ij]}$, $K_i$, and $P_i$, while that of 
time translation is $H$.  Taking natural units, the mass dimensions 
of these generators are $[J_{[ij]}] = [K_i] = 0$ and $[P_i] = [H] = 1$, 
while that of Newton's constant is $[G_N] = -d+2$, where $d$ is the number 
of spacetime dimensions.  It is then easy to find that for $d \geq 4$, 
where gravity is dynamical, there is no term one can add to the commutators 
in Eq.~(\ref{eq:Poincare-alg}) that is linear in generators and has a 
positive integer power of $G_N$.%
\footnote{For $d=3$, one can add terms $\varDelta [J, K_i] = i\gamma 
 G_N P_i$ and $\varDelta [K_1,K_2] = -i\gamma G_N H$, where $J \equiv 
 J_{[12]}$ and $\gamma$ is a real constant, without violating Jacobi 
 identities.  The significance of this is not clear.}
The algebra for the reference frame changes and time translation, 
therefore, is the same as that of the Poincar\'{e} transformations 
in Eq.~(\ref{eq:Poincare-alg}).  The effect of nonzero $G_N$ appears 
as the reduction of the Hilbert space, but not in the transformation 
generators of the Poincar\'{e} group.

\subsection{Selecting the multiverse state}
\label{subsec:selec}

Let us now require that predictions do not depend on the reference frame 
one chooses to describe the multiverse (Hypothesis~II).  Physically, 
this implies that there is neither absolute center nor the frame of 
absolute rest in the multiverse.

Formally, our requirement can be stated as follows.  Suppose we want 
to make physical predictions using projection operators ${\cal O}_X$, 
e.g.\ $X = A$, $A \cap B$, and so on.  The relevant matrix elements 
are then $\bra{\Psi(t)} {\cal O}_X \ket{\Psi(t)}$.  Now, consider 
a multiverse state as viewed from a different reference frame:\ 
$\ket{\Psi'(t)} = S \ket{\Psi(t)}$, where $S$ is the unitary operator 
representing the corresponding reference frame change.  Our requirement 
is then
\begin{equation}
  \bra{\Psi(t)} {\cal O}_X \ket{\Psi(t)} 
  = \bra{\Psi'(t)} {\cal O}_X \ket{\Psi'(t)}
\label{eq:req}
\end{equation}
{\it for arbitrary $S$ and ${\cal O}_X$}.  Note that the operator in 
the right-hand side is {\it not} ${\cal O}'_X = S {\cal O}_X S^\dagger$, 
but the same ${\cal O}_X$ as in the left-hand side.  This equation, 
therefore, has a nontrivial physical content, imposing constraints on 
the multiverse state.  (If we had ${\cal O}'_X$ in the right-hand side, 
then the equation would simply represent a basis change, and thus would 
be trivial.)

In order to satisfy Eq.~(\ref{eq:req}), the multiverse state must satisfy 
$S \ket{\Psi(t)} \propto \ket{\Psi(t)}$, so that it must be a simultaneous 
eigenstate of operators $J_{[ij]}$, $K_i$ and $P_i$.%
\footnote{It is, in principle, possible that the predictions are reference 
 frame independent because the multiverse is in an intrinsically mixed 
 state that satisfies $S \rho(t) S^\dagger = \rho(t)$ at all $t$ but 
 each component $\ket{\psi_i(t)}$ in $\rho(t)$ is {\it not} a simultaneous 
 eigenstate of all the $S$'s.  This is, however, the case only if 
 $\rho(t)$ is the maximally mixed state in ${\cal H}_{\rm QG}$ (because 
 of Schur's lemma), which is observationally excluded as we saw in 
 Section~\ref{subsec:max-ig}.  We must therefore require that each 
 pure-state component leads to reference-frame independent predictions 
 even if the multiverse is in a mixed state.}
One can then easily see from Eq.~(\ref{eq:Poincare-alg}) that this requires 
that the multiverse state is also an eigenstate of $H$, and that the 
eigenvalues under $J_{[ij]}$, $K_i$, $P_i$, and $H$ are all zero.  The 
fact that the multiverse state is an eigenstate of $H$ with zero eigenvalue 
means that
\begin{equation}
  \frac{d}{dt}\ket{\Psi(t)} = 0,
\label{eq:d-dt_Psi}
\end{equation}
i.e.\ the multiverse state is static!  We can therefore write it simply 
as $\ket{\Psi} \equiv \ket{\Psi(t)} = \ket{\Psi(0)}$.  The conditions 
coming from Hypothesis~II can then be summarized as
\begin{equation}
  J_{[ij]} \ket{\Psi} = K_i \ket{\Psi} = P_i \ket{\Psi} = H \ket{\Psi} = 0.
\label{eq:Psi-cond}
\end{equation}
This provides selection conditions for the multiverse state.

In fact, given Hypothesis~I, the conditions in Eq.~(\ref{eq:Psi-cond}) 
follow from a standard procedure of quantizing a system with 
redundancies~\cite{Dirac-book}, if we assume that the multiverse state 
is invariant under the action of global energy and boost operators. 
In this procedure, any gauge redundancy, including general coordinate 
transformations, appears as a supplementary condition imposed on quantum 
states, which eliminates unphysical degrees of freedom from the states. 
Starting from a consistent, general covariant quantum theory of gravity 
(which is presumably string theory), the states are subject to a huge 
number of supplementary conditions, some of which will be used to 
reduce the number of degrees of freedom from that implied by local 
field theory to that suggested by the holographic principle, as in 
Eq.~(\ref{eq:H_M-dimension}).  (This implies that the number of constraints 
is much larger than that of the standard constraints associated with 
classical general coordinate transformations~\cite{DeWitt:1967yk}.) 
In this bigger (more redundant) picture, the framework of 
Refs.~\cite{Nomura:2011dt,Nomura:2011rb} corresponds to the 
scheme in which all the gauge redundancies are explicitly fixed, 
except for the ones associated with the reference frame changes. 
These residual redundancies, i.e.\ those of the reference frame 
changes, must then have their own supplementary conditions imposed 
on the states living in ${\cal H}_{\rm QG}$.

To illustrate this in a simple example, let us consider a spacetime 
that admits rectilinear coordinates $x_i$ in a constant $t$ hypersurface. 
In terms of Hamiltonian and momentum densities, ${\cal H}(x)$ and 
${\cal P}_i(x)$, the Hilbert space ${\cal H}_{\rm QG}$ then corresponds 
to the space of states in which the constraints of the form
\begin{eqnarray}
  && \int\! x_i x_j {\cal H}(x)\, d^3x \ket{\Psi} 
  = \int\! x_i x_j x_k {\cal H}(x)\, d^3x \ket{\Psi} = \cdots 
\nonumber\\
  && \qquad = \int\! x_i x_j {\cal P}_k (x)\, d^3x \ket{\Psi} 
  = \int\! x_i x_j x_k {\cal P}_l(x)\, d^3x \ket{\Psi} = \cdots = 0,
\label{eq:const-1}
\end{eqnarray}
as well as those associated with holography and complementarity, 
are {\it already} imposed; namely, the states in ${\cal H}_{\rm QG}$ 
satisfy these constraints by construction.  On the other hand, the 
constraints of the form
\begin{equation}
  \int\! {\cal H}(x)\, d^3x \ket{\Psi} 
  = \int\! x_i {\cal H}(x)\, d^3x \ket{\Psi} 
  = \int\! {\cal P}_i (x)\, d^3x \ket{\Psi} 
  = \int\! x_i {\cal P}_j(x)\, d^3x \ket{\Psi} = 0
\label{eq:const-2}
\end{equation}
are {\it not} imposed to obtain ${\cal H}_{\rm QG}$, so they must still 
be imposed on the states in ${\cal H}_{\rm QG}$.  Now, the generators 
of time translation and the reference frame changes are given by
\begin{equation}
\begin{array}{c}
  H = \int\! {\cal H}(x)\, d^3x + \epsilon,
\qquad
  P_i = \int\! {\cal P}_i(x)\, d^3x + p_i,
\\[10pt]
  K_i = \int\! x_i {\cal H}(x)\, d^3x + k_i,
\qquad
  J_{[ij]} = \int\! (x_i {\cal P}_j(x) - x_j {\cal P}_i(x)) d^3x + j_{[ij]},
\end{array}
\end{equation}
where we have included global energy $\epsilon$ and momentum $p_i$ 
operators (and the corresponding quantities in $K_i$ and $J_{[ij]}$) 
that represent possible contributions from surface terms.  Such 
terms can indeed arise in asymptotically Minkowski space, and play 
the role of what we consider the total energy and momentum of the 
system~\cite{Arnowitt:1962hi}.

Note that it is the effect of global energy $\epsilon$ that allows 
for any evolution of states in $t$ in quantum gravity, because
\begin{equation}
  \ket{\Psi(t_1)} = e^{-iH(t_1-t_2)} \ket{\Psi(t_2)} 
  = e^{-i\epsilon (t_1-t_2)} \ket{\Psi(t_2)},
\label{eq:gen-evol}
\end{equation}
so unless $\ket{\Psi(t)}$ is a superposition of terms that give different 
values of $\epsilon$, the state is stationary.  In this picture, our 
Hypothesis~II corresponds to the assumption that the multiverse is an 
eigenstate of $\epsilon$ and $k_i$ with vanishing eigenvalues:
\begin{equation}
  \epsilon \ket{\Psi} = k_i \ket{\Psi} = 0,
\label{eq:zero-ep}
\end{equation}
in which case we immediately see that $\ket{\Psi}$ also has zero 
eigenvalues under $p_i$ and $j_{[ij]}$, and that Eq.~(\ref{eq:Psi-cond}) 
follows from the constraints in Eq.~(\ref{eq:const-2}) (and vice versa). 
An important point is that for a state in ${\cal H}_{\rm QG}$, the 
surface terms reside on the (stretched) apparent horizon, so that 
Eq.~(\ref{eq:zero-ep}) is the assumption about the structure of the 
theory on this surface.  This is in the intrinsically quantum gravitational 
regime, over which we currently do not have good theoretical control.

The selection of possible multiverse states, therefore, is boiled down to 
solving the infinite-dimensional matrix equations in Eq.~(\ref{eq:Psi-cond}). 
Here, we assume that there is no other selection condition, i.e.\ 
Eq.~(\ref{eq:Psi-cond}) is enough to fully select the system.  (We 
assume that other supplementary conditions, e.g.\ those associated 
with standard gauge symmetries, are already taken care of.  Also, since 
all the redundancies associated with gravity other than those corresponding 
to the reference frame changes are supposed to be fixed in the present 
framework~\cite{Nomura:2011dt}, there are no more conditions arising 
from considerations of gravity.)  We look for solutions to 
Eq.~(\ref{eq:Psi-cond}) of the form
\begin{equation}
  \ket{\Psi} = \sum_i c_i \ket{\alpha_i},
\qquad
  \sum_i |c_i|^2 < \infty,
\label{eq:sol-cond}
\end{equation}
where $\ket{\alpha_i}$ represents a complete, orthonormal basis in 
${\cal H}_{\rm QG}$, so that the sums of $i$ run to infinity; see 
Eq.~(\ref{eq:dim-inf}).  The normalizability condition here is imposed 
for the following (usual) reason.  Suppose there are normalizable solutions 
$\ket{\Psi_I}$ ($I = 1,\cdots,N$) satisfying Eq.~(\ref{eq:sol-cond}), as 
well as non-normalizable solutions $\ket{\Psi_I}$ ($I = N+1,\cdots,K$). 
The non-normalizable solutions will have coefficients which strongly 
diverge as the dimensions of corresponding Hilbert subspaces 
${\cal H}_{\cal M}$ become large.  This is because the process transforming 
an element of ${\cal H}_{\cal M}$ to that of ${\cal H}_{\cal M'}$ with 
${\rm dim}\,{\cal H}_{\cal M'} < {\rm dim}\,{\cal H}_{\cal M}$ becomes 
highly suppressed as ${\rm dim}\,{\cal H}_{\cal M}$ gets large (because 
of Eq.~(\ref{eq:dim-inf})).  Let us now imagine regulating the sums of 
$i$ as $\sum_i \rightarrow \sum_{i=1}^{n}$, in which case we can normalize 
all the solutions so that $\inner{\Psi_I}{\Psi_J} = \delta_{IJ}$ for 
$I,J = 1,\cdots,K$.  We can then consider state $\rho = \sum_{I,J=1}^{K} 
d_{IJ} \ket{\Psi_I} \bra{\Psi_J}$ with arbitrary finite positive 
semi-definite Hermitian matrix $d_{IJ}$, and calculate probabilities 
arising from $\rho$ using a projection operator that selects (a finite 
number of) configurations compatible with some condition $X$:\ ${\cal O}_X 
= \sum_{i \in X} \ket{\alpha_i} \bra{\alpha_i}$.  The resulting 
probabilities are the same as those arising from $\rho' = \sum_{I,J=1}^{N} 
d_{IJ} \ket{\Psi_I} \bra{\Psi_J}$, i.e.\ the state obtained by eliminating 
all the non-normalizable solutions from $\rho$, up to terms disappearing 
for $n \rightarrow \infty$.  Therefore, the non-normalizable solutions 
can all be dropped from physical considerations.

\begin{figure}[t]
\begin{center}
  \includegraphics[width=17cm]{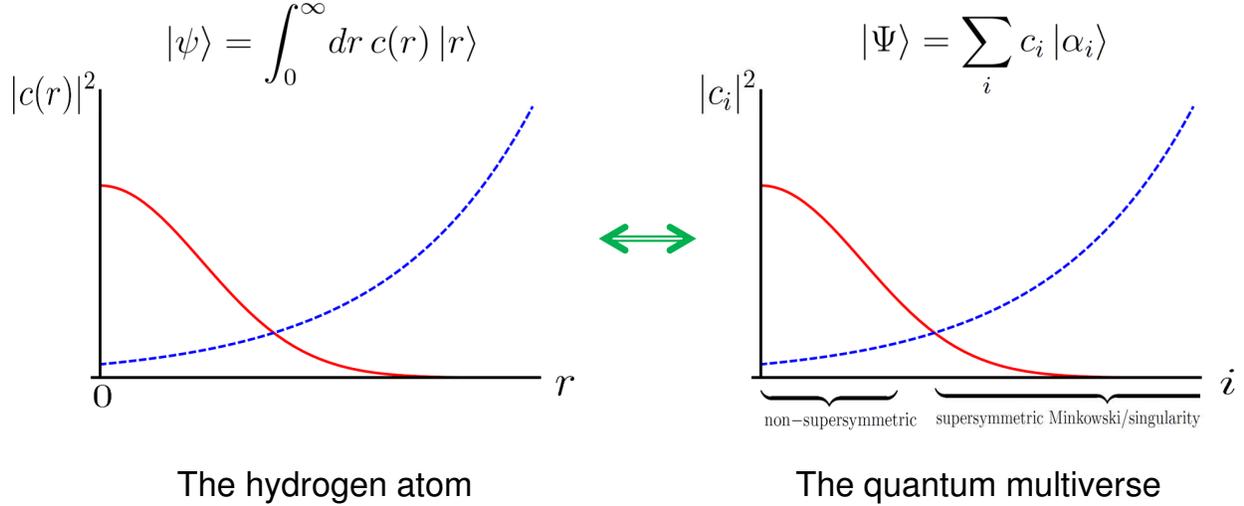}
\caption{A schematic depiction of the analogy between the hydrogen 
 atom and the quantum multiverse.  In the case of the hydrogen atom, 
 the only relevant states are those that satisfy the Schr\"{o}dinger 
 equation and are normalizable in the Hilbert space spanned by $\ket{r}$ 
 (solid line); the non-normalizable modes are irrelevant (dashed 
 line).  In the quantum multiverse, the relevant states are those 
 that satisfy Eq.~(\ref{eq:Psi-cond}) and are normalizable in Hilbert 
 space ${\cal H}_{\rm QG}$ (solid line); the non-normalizable modes, 
 which have diverging coefficients for supersymmetric Minkowski or 
 singularity states, are irrelevant (dashed line).}
\label{fig:analogy}
\end{center}
\end{figure}

The Hilbert space relevant for the multiverse ${\cal H}_{\rm Multiverse}$, 
then, is spanned by the normalizable solutions to Eq.~(\ref{eq:Psi-cond}), 
and so is much smaller than ${\cal H}_{\rm QG}$:
\begin{equation}
  {\cal H}_{\rm Multiverse} \subset {\cal H}_{\rm QG},
\qquad
  {\rm dim}\,{\cal H}_{\rm Multiverse} \ll {\rm dim}\,{\cal H}_{\rm QG}.
\label{eq:H_multiverse}
\end{equation}
We note that this situation is analogous to usual quantum mechanical 
systems, e.g.\ a hydrogen atom.  In the hydrogen atom, the state factor 
corresponding to a radial wavefunction $c(r)$ can be written as $\ket{\psi} 
= \int_0^\infty\!dr\, c(r) \ket{r}$.  The only states relevant to physics 
of the hydrogen atom are those satisfying the normalizability condition 
$\int_0^\infty\!dr\, |c(r)|^2 < \infty$ in the Hilbert space spanned 
by $\ket{r}$.  The other, non-normalizable solutions (which behave as 
$\ln c(r) \sim r$ at large $r$) are irrelevant.  The situation in the 
quantum multiverse is similar.  The non-normalizable solutions have 
infinitely strong supports in supersymmetric Minkowski vacua or singularity 
worlds, which have infinite-dimensional Hilbert spaces.  These solutions, 
therefore, are irrelevant in making predictions in a ``realistic world,'' 
i.e.\ in a universe that has nonzero free energy.  The only relevant states 
are those that are normalizable in the Hilbert space of quantum gravity, 
${\cal H}_{\rm QG}$.  For a schematic drawing of this analogy, see 
Fig.~\ref{fig:analogy}.

\subsection{The static multiverse states in {\boldmath ${\cal H}_{\rm QG}$}}
\label{subsec:static}

We now discuss how our conditions Eqs.~(\ref{eq:d-dt_Psi},~\ref{eq:sol-cond}) 
can be compatible with Eq.~(\ref{eq:asympt}), which says that a generic 
multiverse state in ${\cal H}_{\rm QG}$ will evolve into a superposition 
of supersymmetric Minkowski and singularity states as $t \rightarrow 
\infty$.  In order for Eq.~(\ref{eq:d-dt_Psi}) to be satisfied, the 
coefficients $c_i$ of all the terms in $\ket{\Psi(t)} = \ket{\Psi}$ 
must be constant when expanded in components $\ket{\alpha_i}$.  In 
a basis in which $\ket{\alpha_i}$ in ${\cal H}$ have well-defined 
semi-classical configurations, the evolution operator $\exp(-iHt)$ 
(and thus $H$ as well) is not diagonal.  Therefore, the processes in 
Eq.~(\ref{eq:asympt}) will occur for {\it generic} $\ket{\Psi(t)}$, 
but they must exactly be canceled by some ``inverse processes'' in 
$\ket{\Psi}$.  In particular, in order for the normalization condition 
in Eq.~(\ref{eq:sol-cond}) to be satisfied, this must occur before 
the state is dissipated into infinite-dimensional Hilbert space.

Let us consider a physical configuration in a Minkowski universe in which 
there is a bubble wall surrounding us, which, however, is contracting 
toward us rather than expanding away.  Such a configuration, which is 
exactly the time reversal of a usual expanding bubble configuration, is 
physically allowed, as the fundamental equation of the theory is symmetric 
under $t \rightarrow -t$.  Usually, we do not consider this kind of 
configuration as it is only an exponentially small subset of all 
the configurations allowed by the theory; in particular, there is only 
an exponentially small probability for forming such a configuration 
starting from a generic, e.g.\ thermal, state.  We are, however, 
now considering very special states, i.e.\ the states that satisfy 
Eqs.~(\ref{eq:d-dt_Psi},~\ref{eq:sol-cond}), and in these states such 
configurations could balance the ``loss'' of semi-classically unstable 
states in Eq.~(\ref{eq:asympt}).  For example, the entire multiverse 
state is so ``fine-tuned'' that a reheating that occurs in a Minkowski 
universe produces exactly the configuration that puts the system back 
to (a superposition of) states in unstable vacua.  Similar processes 
must also occur for singularities.  Note that since these processes are 
exponentially suppressed under normal circumstances, they are invisible 
in the usual semi-classical analysis.

The states given by Eq.~(\ref{eq:sol-cond}) are the ones in which all 
these and other processes are balanced.%
\footnote{There is also the possibility that some (or all) of the 
 states given by Eq.~(\ref{eq:sol-cond}) do not contain any Minkowski 
 or singularity components.  This does not affect any of our discussions 
 below.}
Since the inverse processes are unlikely to occur at the zero density, 
these states will explore only a finite-dimensional portion of Minkowski 
vacua (see the discussion at the end of Section~\ref{subsec:Hilbert}). 
The number of independent states, therefore, may well be finite:\ 
${\rm dim}\,{\cal H}_{\rm Multiverse} < \infty$, which we will assume 
to be the case.  Note that the sizes of various elements in $H$ represented 
as a matrix acting on ${\cal H}_{\rm QG}$ differ significantly; in fact, 
they are expected to differ exponentially, or even double-exponentially, 
as some of the processes are highly suppressed.  This implies that the 
values of $|c_i|$'s in Eq.~(\ref{eq:sol-cond}) will also vary significantly. 
The resulting states $\ket{\Psi}$, therefore, are not excluded by 
corollary~($\ast$) in Section~\ref{subsec:arrow-time}.  The structure 
of $\ket{\Psi}$, and its consistency with observation, will be discussed 
further in Section~\ref{sec:consistency}.

\subsection{Predictions in the static quantum multiverse}
\label{subsec:pred}

The number of independent normalizable solutions to Eq.~(\ref{eq:Psi-cond}) 
will depend on the structure of the multiverse, i.e.\ issue~(a) in 
Section~\ref{subsec:s-c}.  In particular, the existence of a solution 
requires $H$ to take a certain special form (so that it has at least 
one normalizable, zero-eigenvalue eigenvector), which we assume to be 
the case.  Suppose there are $N$ such solutions $\ket{\Psi_I}$ ($I = 
1,\cdots,N = {\rm dim}\,{\cal H}_{\rm Multiverse} < \infty$).  How can 
the physical predictions be made?

If $N = 1$, the multiverse state is simply $\ket{\Psi} \equiv 
\ket{\Psi_1}$.  The probabilities are then given by the generalized 
Born rule, Eq.~(\ref{eq:prob}), but now without the $t$ integrals. 
(They simply give a constant factor $\int_{-\infty}^{+\infty} dt$, 
which cancels between the numerator and denominator.)  The final 
formula is given by Eq.~(\ref{eq:prob-final}), which we reproduce here:
\begin{equation}
  P(B|A) = \frac{\bra{\Psi} {\cal O}_{A \cap B} \ket{\Psi}}
    {\bra{\Psi} {\cal O}_A \ket{\Psi}}.
\nonumber
\end{equation}
As discussed in Section~\ref{subsec:evolve}, this formula can be used 
to answer any physical questions, including those about a system that 
we view as dynamically evolving.

In the case that $N > 1$, any multiverse states of the form $\ket{\Psi} 
= \sum_{I=1}^N c_I \ket{\Psi_I}$ or $\rho = \sum_{I,J=1}^{N} d_{IJ} 
\ket{\Psi_I} \bra{\Psi_J}$ are allowed.  In the absence of more information 
(or selection conditions), it is natural to assume that the multiverse 
is in the maximally mixed state
\begin{equation}
  \rho = \frac{1}{N} \sum_{I=1}^N \ket{\Psi_I} \bra{\Psi_I},
\label{eq:multiverse-mixed}
\end{equation}
where we have taken $\ket{\Psi_I}$'s to be orthonormal.  This state 
is invariant under the basis change $\ket{\Psi_I} \rightarrow U_{IJ} 
\ket{\Psi_J}$, and is reduced to $\ket{\Psi} = \ket{\Psi_1}$ for 
$N = 1$.  The probabilities are given by the mixed-state version 
of Eq.~(\ref{eq:prob-final}):
\begin{equation}
  P(B|A) = \frac{{\rm Tr}\left[ \rho\, {\cal O}_{A \cap B} \right]}
    {{\rm Tr}\left[ \rho\, {\cal O}_A \right]}.
\label{eq:prob-final-mixed}
\end{equation}
Note that Eq.~(\ref{eq:multiverse-mixed}), i.e.\ the maximally mixed state 
in ${\cal H}_{\rm Multiverse}$, is different from Eq.~(\ref{eq:rho-MM}), 
i.e.\ the maximally mixed state in ${\cal H}_{\rm QG}$, in which the 
sum runs over all the possible states in ${\cal H}_{\rm QG}$ including 
the ones that do not satisfy Eq.~(\ref{eq:Psi-cond}).  The state 
in Eq.~(\ref{eq:multiverse-mixed}), therefore, is not excluded by 
corollary~($\ast$) in Section~\ref{subsec:arrow-time}.

\section{Consistency with Observation}
\label{sec:consistency}

In this section we discuss the consistency of the present scenario with 
observation, specifically the observed arrow of time.  Our approach 
here will be to allow for making assumptions on the structures of $H$ 
and ${\cal H}_{\rm QG}$ (unless they are inconsistent with what we 
already know about string theory), and to see if the scenario is 
consistent.  We do not claim that all of these assumptions are absolutely 
necessary---our purpose here is to argue that, despite its naive 
appearance, the scenario is not excluded by observation.  More detailed 
analysis/modeling of the landscape will be left for future work.

\subsection{The structure of {\boldmath ${\cal H}_{\rm QG}$}}
\label{subsec:structure}

Solutions to Eq.~(\ref{eq:Psi-cond}) depend on the structure of 
${\cal H}_{\rm QG}$ as well as the form of $H$ (and other operators). 
Here we assume that ${\cal H}_{\rm QG}$ contains only ``cosmologically 
relevant'' states.  The minimally required set of ${\cal H}_{\cal M}$'s 
that must be included in ${\cal H}$, i.e.\ in the right-hand side of 
Eq.~(\ref{eq:ST-H}), will then be those of FRW universes corresponding 
to all the possible vacua in the theory (and their straightforward 
generalizations, e.g.\ those of FRW universes with black holes).  Not 
all spacetime must be contained in ${\cal H}$; for example, ${\cal H}$ 
need not contain a stable anti-de~Sitter space without a singularity, 
which might only be a mathematical idealization because it does not 
arise through dynamical evolution in the FRW universes.

For each vacuum $I$ of the theory, the number of states associated with 
an FRW universe in $I$ is estimated as
\begin{equation}
  {\cal N}_I \,\,= 
    \sum_{n=\exp({\cal A}_{I,{\rm min}}/2)}^{\exp({\cal A}_{I,{\rm max}}/2)}
    \!\!\!\! n
  \,\,\simeq\,\,
     \frac{1}{2}\, e^{{\cal A}_{I,{\rm max}}},
\label{eq:N_I}
\end{equation}
where ${\cal A}_{I,{\rm min}}$ and ${\cal A}_{I,{\rm max}}$ are the minimum 
and maximum areas of the apparent horizon in this universe, and we have 
used ${\cal A}_{I,{\rm max}} \gg {\cal A}_{I,{\rm min}}$ in the last 
equation.  While possible deformations of the apparent horizon, e.g.\ 
by the existence of black holes, can have corrections to the explicit 
expression, we expect that the above estimate gives a qualitatively 
correct result:\ $\ln {\cal N}_I \approx O({\cal A}_{I,{\rm max}})$. 
The area ${\cal A}_{I,{\rm max}}$ is given by the inverse of the absolute 
value of the vacuum energy density (in Planck units) ${\cal A}_{I,{\rm max}} 
\sim 1/|\rho_{\Lambda,I}|$, since in a de~Sitter universe the apparent 
horizon approaches the event horizon at late times, while in an 
anti-de~Sitter universe it has the maximum area when $p$ hits the 
singularity at $t \sim 1/|\rho_{\Lambda,I}|^{1/2}$.  We therefore find
\begin{equation}
  \ln {\cal N}_I \sim \frac{1}{|\rho_{\Lambda,I}|}.
\label{eq:ln-N_I}
\end{equation}
This implies that the number of states associated with a vacuum with 
$\rho_{\Lambda,I} \neq 0$ is finite.

\subsection{The arrow of time in the static multiverse}
\label{subsec:arrow}

We now consider a solution to the equation $H \ket{\Psi} = 0$, a part of 
the conditions in Eq.~(\ref{eq:Psi-cond}).  We can view this equation as 
requiring that $\ket{\Psi}$ is in a stationary state in ${\cal H}_{\rm QG}$. 
(In fact, the equation is stronger than that, since the eigenvalue of 
$H$ must be zero.)  In particular, it implies that the probability current 
creating states in vacuum $I$ must be balanced with that destroying those 
for each $I$ (in fact, each state in $I$).  At the semi-classical level, 
this condition is impossible to satisfy for terminal vacua.  As discussed 
in Section~\ref{subsec:static}, however, our state is special, obtained 
after solving the ``quantization condition'' $H \ket{\Psi} = 0$, so that 
it can also be satisfied for these vacua.

Let us now consider vacuum $J$ that can support any observer, either 
an ordinary observer or a Boltzmann brain.  We will argue that the arrow 
of time is predicted if the following three conditions are met for 
all possible $J$'s:
\begin{itemize}
\item[(I)]
Transitions to states in $J$ from those in other vacua are mainly 
through the states having low coarse-grained entropies in $J$, i.e.\ 
elements of ${\cal H}_{\cal M}$ with $\ln {\rm dim}\,{\cal H}_{\cal M} 
\ll {\cal A}_{J,{\rm max}}$.
\item[(II)]
Subsequent evolution in vacuum $J$ produces ordinary observers 
with probability $\epsilon_J$, which may be suppressed exponentially 
but not double-exponentially.
\item[(III)]
The rate of producing Boltzmann brains $\Gamma_{{\rm BB},J}$ in vacuum $J$, 
which is double-exponentially suppressed (see, e.g.~\cite{Bousso:2008hz}), 
is smaller than the decay rate $\Gamma_J$ of the vacuum itself.
\end{itemize}
Namely, if the structure of $H$ is such that it satisfies all these 
conditions, then the scenario is compatible with observation.  (The 
``transitions'' and ``evolution'' here, of course, refer to the apparent 
ones in $\ket{\Psi}$, which is in itself static.)

To see this, let us consider the distribution of the size of the 
coefficients $|c^J_i|$ of various terms in $\ket{\Psi}$ corresponding 
to the states in vacuum $J$, $\ket{\alpha^J_i}$.  For this purpose, 
we define the quantity $P^J_\tau$ corresponding to the probability 
for a universe to be at FRW time $t_{\rm FRW}$ between $\tau$ and 
$\tau+d\tau$:
\begin{equation}
  P^J_\tau\, d\tau 
  = \sum_{i | \tau < t_{\rm FRW} < \tau+d\tau}\!\!\!\! |c^J_i|^2,
\end{equation}
where $t_{\rm FRW}$ should be specified by physical configurations 
in $\ket{\alpha^J_i}$.  The distribution of $P^J_\tau$ then follows 
from the definition of $\Gamma_J$:
\begin{equation}
  P^J_\tau = P^J_0 e^{-\Gamma_J\, \tau},
\label{eq:PJtau}
\end{equation}
where we have assumed that the transitions to states in $J$ occur 
at $\tau = 0$ either through Coleman-De~Luccia~\cite{Coleman:1980aw} 
or Hawking-Moss~\cite{Hawking:1981fz} processes (or their inverses), 
although our conclusion is insensitive to this assumption.  Note that 
in these cases it is indeed natural to expect that states just after 
the transitions are the ones having low coarse-grained entropies, 
i.e.\ in ${\cal H}_{\cal M}$ with $\ln {\rm dim}\,{\cal H}_{\cal M} 
\ll {\cal A}_{J,{\rm max}}$, because both the start and end points of 
the Coleman-De~Luccia tunneling in field space are away from local minima 
(if the false vacuum has a positive vacuum energy), and the Hawking-Moss 
transition is a thermal process occurring through the field climbing 
up the potential barrier~\cite{Weinberg:2006pc}.

Now, the definitions of $\epsilon_J$ and $\Gamma_{{\rm BB},J}$ in 
(II) and (III) above imply that if we compute the probability of 
$\ket{\Psi}$ containing ordinary observers (OO) or Boltzmann brains 
(BB) in vacuum $J$ using the corresponding projection operators 
${\cal O}_{{\rm OO},J}$ and ${\cal O}_{{\rm BB},J}$, then we obtain
\begin{eqnarray}
  && \bra{\Psi} {\cal O}_{{\rm OO},J} \ket{\Psi} \sim \epsilon_J P^J_0,
\\
  && \bra{\Psi} {\cal O}_{{\rm BB},J} \ket{\Psi} 
  \sim \Gamma_{{\rm BB},J} \int\! P^J_\tau\, d\tau 
  = \frac{\Gamma_{{\rm BB},J}}{\Gamma_J} P^J_0.
\label{eq:OO-BB}
\end{eqnarray}
Here, the projection operators select observers in a specific range 
of location and velocity with respect to $p$, although the results 
do not depend on the chosen location or velocity because of 
Eq.~(\ref{eq:Psi-cond}).  Under conditions~(II) and (III), 
this gives
\begin{equation}
  \frac{\bra{\Psi} {\cal O}_{{\rm BB},J} \ket{\Psi}}{\bra{\Psi} 
    {\cal O}_{{\rm OO},J} \ket{\Psi}} 
  \sim \frac{\Gamma_{{\rm BB},J}}{\epsilon_J \Gamma_J} \lll 1,
\label{eq:BB-OO-ratio}
\end{equation}
where we have used the fact that $\Gamma_{{\rm BB},J}$ is 
double-exponentially suppressed while $\epsilon_J$ is not.  (In fact, 
we only need $\epsilon_J > \Gamma_{{\rm BB},J}/\Gamma_J$ to obtain this 
result, so $\epsilon_J$ may be double-exponentially suppressed.)  We 
therefore find that the overwhelming majority of observers are indeed 
ordinary observers, and thus perceive time's arrow (as discussed in 
Section~\ref{subsec:max-ig}).

Perhaps not surprisingly, the conditions described above are 
similar to the ones obtained in Ref.~\cite{Bousso:2011aa} in the 
context of the evolving multiverse, despite the fact that the overall 
physical pictures are rather different.  One distinct feature of the 
present scenario in this respect is that since there is no ``initial 
vacuum,'' the absolute nonexistence of Boltzmann brains in such a vacuum 
($\Gamma_{{\rm BB},\ast} = 0$ in the notation of Ref.~\cite{Bousso:2011aa}) 
need not be imposed.  In any case, as discussed in Ref.~\cite{Bousso:2011aa}, 
the conditions described above, in particular (I), are likely to be 
satisfied in the string landscape.  It is, therefore, quite promising 
that the scenario discussed in this paper is indeed consistent with 
observation in the realistic string theory setup.

\section{Discussions}
\label{sec:discuss}

In this paper we have studied the multiverse in the quantum mechanical 
framework recently proposed in Refs.~\cite{Nomura:2011dt,Nomura:2011rb}. 
By requiring that the laws of quantum mechanics are not violated 
(Hypothesis~I) and that physical predictions do not depend on the 
reference frame one chooses to described the multiverse (Hypothesis~II), 
we have found that the multiverse state must be static; in particular, 
the multiverse does not have a beginning or end.

Despite its naive appearance, the scenario does not contradict 
observation, including the fact that we observe that time flows in 
a definite direction.  {\it The arrow of time is simply an emergent 
phenomenon that is occurring in the branch (terms) corresponding to 
our universe in the static multiverse state}---the terms that would 
be obtained by evolving the system from lower entropy states have much 
larger coefficients than the terms that cannot.  The scenario is summarized 
by the selection conditions in Eq.~(\ref{eq:Psi-cond}), imposed on the 
states in ${\cal H}_{\rm QG}$.  With these conditions, any multiversal 
questions can be answered using the Born rule, Eq.~(\ref{eq:prob-final}) 
or (\ref{eq:prob-final-mixed}), {\it without any additional input}, once 
the explicit form of the operators such as $H$ is known.  This scenario, 
therefore, provides a completion of the framework of the quantum 
multiverse in Refs.~\cite{Nomura:2011dt,Nomura:2011rb}.

The supplementary condition of the form $H \ket{\Psi} = 0$ has certainly 
been considered before---indeed, this is nothing but the well-known 
Wheeler-DeWitt equation~\cite{DeWitt:1967yk}.  The scenario presented 
here, however, differs from standard applications of this equation 
in several important ways:
\begin{itemize}
\item
The redundancies associated with gravity are much larger than what are 
usually imagined.  In particular, they reduce the Hilbert space in such 
a way that it contains only the spacetime region within the reference 
point's (stretched) apparent horizon~\cite{Nomura:2011dt,Nomura:2011rb}. 
This is important to avoid ambiguities associated with eternally 
inflating spacetime.  The ultimate origin of these large redundancies 
will, presumably, be string theory.
\item
We apply the supplementary conditions corresponding to the whole 
set of time translation and reference frame changes with zero global 
charges, even if the universe is not closed.  Since spacetime is defined 
only within the apparent horizon, this requires the assumption on the 
structure of the theory on this surface, which is intrinsically quantum 
mechanical.  Note that it is this assumption that is responsible for 
the static nature of the multiverse state, which in turn excludes the 
possibility for the multiverse to have a beginning or end.
\item
We analyze the consequences of the supplementary conditions at the 
microscopic level.  This selects very special states that are 
not visible in the analysis at the semi-classical level.  In fact, 
normalizable solutions to the conditions correspond to the states 
in which the processes of Eq.~(\ref{eq:asympt}) are balanced with 
the inverse processes, which put the system back from terminal vacua 
to unstable vacua.
\end{itemize}
It is quite satisfying that such simple requirements as Hypotheses~I 
and II lead to a consistent and predictive scheme for the entire 
multiverse.

Finally, it is instructive to draw a close analogy between the situation 
in the quantum multiverse described here and that in the standard, 
hydrogen atom.  As is well known, the hydrogen atom cannot be correctly 
described using classical mechanics.  Any orbit of the electron is 
unstable with respect to the emission of synchrotron radiation.  Even 
if we artificially ignore the emission, the electron can orbit the 
nucleus at an arbitrary radius, unable to explain the discrete spectral 
lines.  The solution to these problems is intrinsically quantum mechanical, 
i.e.\ quantum mechanics is responsible for the very existence of the 
hydrogen atom, not just providing a correction to the classical picture.

The situation in the quantum multiverse is similar.  At the semi-classical 
level, the multiverse is unstable to the decay to terminal states, 
such as supersymmetric Minkowski vacua and singularities.  Even if 
we artificially ignore the process of vacuum decays, it would lead 
to phenomena such as Poincar\'{e} recurrence, contradicting observation 
(the dominance of Boltzmann brains).  The picture presented here says 
that the solution to these problems is {\it intrinsically quantum 
mechanical}---one cannot see it in the usual semi-classical analysis. 
The multiverse state is very special:\ a normalizable state satisfying 
the ``quantization conditions'' of Eq.~(\ref{eq:Psi-cond}), as in the 
case of the hydrogen atom.  In the case of the hydrogen atom, these 
conditions make the dimension of Hilbert space from continuous infinity 
$\psi(r, \theta, \varphi)$ to countable infinity $(n,l,m)$.  In the 
quantum multiverse, they will presumably make it from countable 
infinity to finite:\ ${\rm dim}\,{\cal H}_{\rm QG} \rightarrow 
{\rm dim}\,{\cal H}_{\rm Multiverse}$.

After all, quantum mechanics treats the multiverse very similarly to 
the hydrogen atom.  Our job is then to figure out the precise structure 
of the multiverse, a system which we are a part of.  Hopefully, further 
progress in string theory will serve this purpose.

\section*{Acknowledgments}

I would like to thank Alan Guth and Grant Larsen for useful conversations. 
This work was supported in part by the Director, Office of Science, Office 
of High Energy and Nuclear Physics, of the US Department of Energy under 
Contract DE-AC02-05CH11231, and in part by the National Science Foundation 
under grant PHY-0855653.

\end{document}